\documentclass[sigconf]{acmart}
\usepackage{amsthm}
\usepackage{subcaption}
\usepackage{paralist}
\usepackage{bm}
\usepackage{amsfonts}
\usepackage{multirow}
\usepackage{cases}
\usepackage{booktabs}
\usepackage{threeparttable}
\usepackage{epstopdf}
\usepackage{upgreek}
\usepackage{endnotes}
\usepackage{etoolbox}
\usepackage{algpseudocode}
\usepackage{amsmath}
\usepackage{algorithm}
\usepackage{balance}
\usepackage{xspace}
\usepackage{array}
\usepackage{enumitem}
\usepackage{hyperref}

\AtBeginDocument{%
  \providecommand\BibTeX{{%
    \normalfont B\kern-0.5em{\scshape i\kern-0.25em b}\kern-0.8em\TeX}}}
\newcommand{\eat}[1]{}
\newcommand{\paratitle}[1]{\vspace{1.5ex}\noindent \textbf{#1}}
\newcommand{\ie}{\emph{i.e.,}\xspace}

\newcommand{\baby}{\textsc{MGNM}\xspace}

\copyrightyear{2022}
\acmYear{2022}
\setcopyright{acmcopyright}\acmConference[SIGIR '22]{Proceedings of the 45th
International ACM SIGIR Conference on Research and Development in Information
Retrieval}{July 11--15, 2022}{Madrid, Spain}
\acmBooktitle{Proceedings of the 45th International ACM SIGIR Conference on
Research and Development in Information Retrieval (SIGIR '22), July 11--15, 2022,
Madrid, Spain}
\acmPrice{15.00}
\acmDOI{10.1145/3477495.3532081}
\acmISBN{978-1-4503-8732-3/22/07}

\begin{document}
\fancyhead{}
\title{When Multi-Level Meets Multi-Interest: A Multi-Grained Neural Model for Sequential Recommendation}\thanks{$^\dagger$Chenliang Li is the corresponding author. Work done when Yu Tian was an intern at Kuaishou.}
\author{Yu Tian$^{1}$, Jianxin Chang$^{2}$, Yanan Niu$^{2}$, Yang Song$^{2}$, Chenliang Li$^{1\dagger}$}
\affiliation{
\institution{
  $^1$Key Laboratory of Aerospace Information Security and Trusted Computing, Ministry of Education, School of Cyber Science and Engineering, Wuhan University, Wuhan, 430072, China \\
  s.braylon1002@gmail.com; cllee@whu.edu.cn\\
  $^2$Kuaishou Technology Co., Ltd., Beijing, 10010, China\\\{changjianxin,niuyanan,yangsong\}@kuaishou.com
  }
  \country{}
}

\def\authors{Yu Tian, Jianxin Chang, Yanan Niu, Yang Song, Chenliang Li}
\renewcommand{\shortauthors}{Yu Tian, et al.}

\begin{abstract}
Sequential recommendation aims at identifying the next item that is preferred by a user based on their behavioral history. Compared to conventional sequential models that leverage attention mechanisms and RNNs, recent efforts mainly follow two directions for improvement: \textit{multi-interest learning} and \textit{graph convolutional aggregation}. Specifically, multi-interest methods such as ComiRec and MIMN, focus on extracting different interests for a user by performing historical item clustering, while graph convolution methods including TGSRec and SURGE
elect to refine user preferences based on multi-level correlations between historical items. Unfortunately, neither of them realizes that these two types of solutions can mutually complement each other, by aggregating multi-level user preference to achieve more precise multi-interest extraction for a better recommendation. 
To this end, in this paper, we propose a unified \textbf{m}ulti-\textbf{g}rained \textbf{n}eural \textbf{m}odel~(named \baby) via a combination of multi-interest learning and graph convolutional aggregation. Concretely, \baby first learns the graph structure and information aggregation paths of the historical items for a user. It then performs graph convolution to derive item representations in an iterative fashion, in which the complex preferences at different levels can be well captured. Afterwards, a novel sequential capsule network is proposed to inject the sequential patterns into the multi-interest extraction process, leading to a more precise interest learning in a multi-grained manner. Experiments on three real-world datasets from different scenarios demonstrate the superiority of \baby against several state-of-the-art baselines. The performance gain over the best baseline is up to 3.12\% and 4.35\% in terms of NDCG@5 and HIT@5 respectively, which is one of the largest gains in recent development of sequential recommendation. Further analysis also demonstrates that \baby is robust and effective at user preference understanding at multi-grained levels.
\end{abstract}

\ccsdesc[500]{Information systems~Recommender systems}

\keywords{Sequential Recommendation, Multi-Interest Learning, Graph Neural Network}

\maketitle
\section{Introduction}


With the rapid development of the Internet, recommender systems have become an important tool to solve information overload and enhance competitiveness for many online services such as news feeds, E-commerce, advertising, and social media. 
Obviously, sequential recommendation, which aims to identify the next item that a user will prefer in terms of her historical behaviors, has drawn increasing attention. The core challenge is how to capture the accurate interests from the user's complex behaviors.

In the past few years, many sequential recommendation solutions have been proposed to model sequential patterns for preference learning. Specifically, earlier works aim to learn a user embedding vector by encoding the user's overall preference from her complex behavior sequence~\cite{tang2018personalized, hidasi2015session, vaswani2017attention, sha2017interpretable, yu2019adaptive}. Typically, a sequence modeling technique is applied over the user behavior sequence. For example, GRU4Rec~\cite{hidasi2015session} uses the GRU module to encode preference signals from user behavior sequences. CASER~\cite{tang2018personalized} considers the sequence of item embeddings as an image and learns sequential patterns via horizontal and vertical convolutional filters.

\begin{figure}[!]
  \centering
  \includegraphics[width=0.38\textwidth]{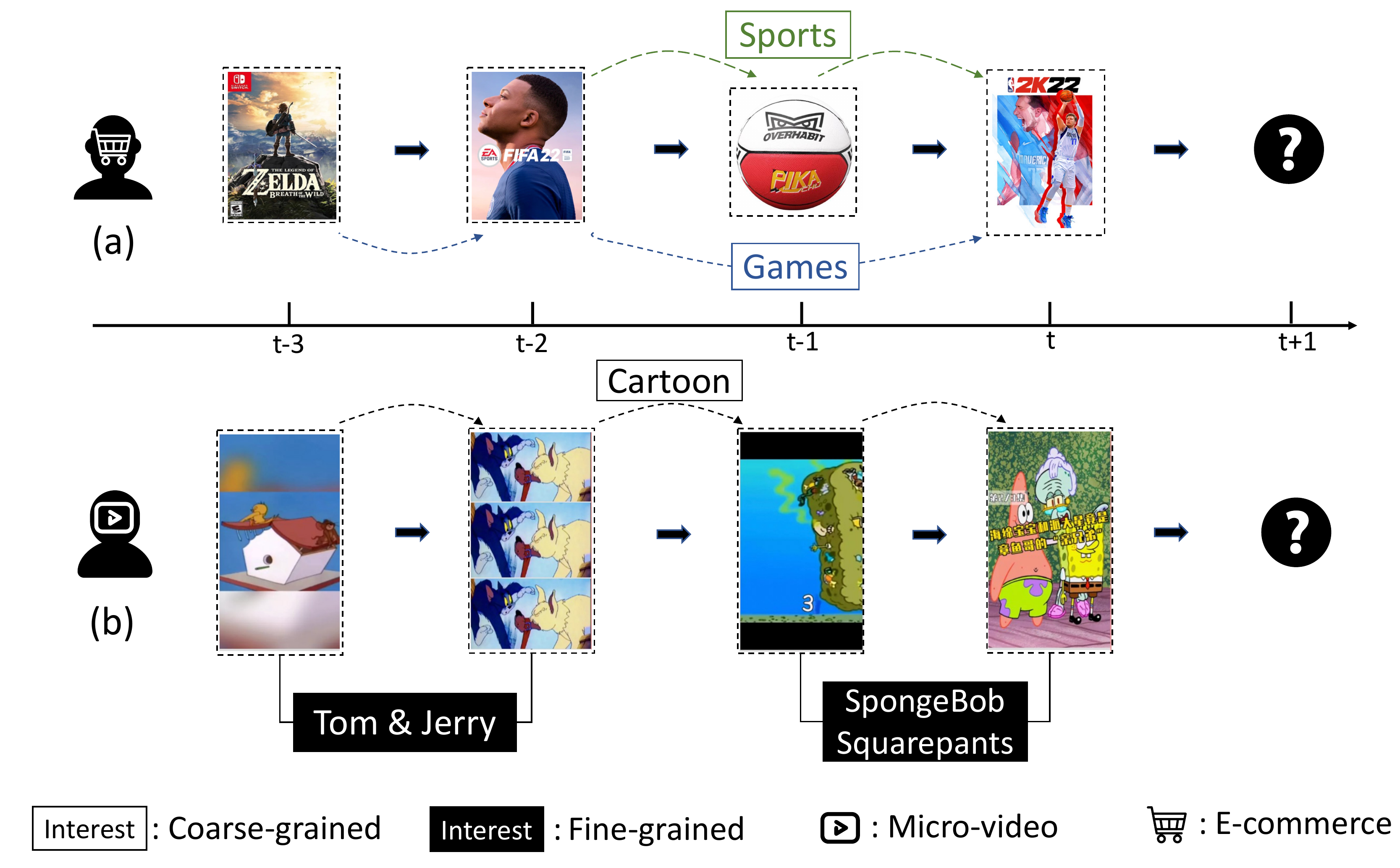}
  \caption{Partial viewing history of two real users in e-commerce and micro-video scenes, respectively. For the user (a), there is a problem that items have an impact on two interests at the same time, i.e. interest overlapping at the $(t-2)$-th and $t$-th timestamps. For the user (b), there are two different levels of interest in her interaction history: coarse-grained (i.e. Cartoon) and fine-grained (i.e. Tom and Jerry).}
  \Description{Partial viewing history of real user in micro video scenes.}
  \label{fig:hist}
\end{figure}

Despite the great success achieved by these solutions, all of them ignore the discrimination of different interests by compositing multifaceted preferences into a single vector.
 Figure~\ref{fig:hist} illustrates the click sequences of two users from the E-commerce and Micro-video datasets, respectively. Here, each video is displayed by its first frame. From Figure~\ref{fig:hist}(a), in this short click history, there are two main interests: sports and games. To address the above problem, a handful of multi-interest solutions are proposed recently. These methods are devised to learn accurate preference vectors for each user by multi-interest modeling. Generally, a multi-interest network is utilized to explicitly encode the multiple interests according to relevant information of items in the behavior sequence. For example, MIMN [16] utilizes memory induction units as multiple channels to derive multiple interests from the user's behavior sequence, which delivers large performance gain in the display advertising system of Alibaba. What's more, MIND~\cite{li2019multi} and ComiRec~\cite{cen2020controllable}  have been improved respectively on the basis of Capsule Network~(CapsNet)~\cite{capsule17}, and the online system has also gained benefits.

All these multi-interest models, however, take the item as the minimum interest modeling unit, lacking the ability of modeling complex, dynamic and high-order user behaviors. More specifically, as shown in Figure~\ref{fig:hist}~(a), the user mainly focuses on sports~(shown in green) and games~(shown in blue). Note that the two items in the $(t-2)$-th and $t$-th timestamps have an impact on the modeling of both two interests~(\ie interest overlapping). In this case, it is difficult to decompose accurately for the existing multi-interest solutions. 
Moreover, Figure~\ref{fig:hist}~(b) shows that a user's interest would be in different granularities. To address this problem, some efforts propose to combine the sequential modeling with graph neural networks~\cite{fan2021continuous, chang2021sequential}. They build an item graph for the historical interacted items and perform the graph convolution to aggregate the user preference in different levels. However, in comparison to multi-interest solutions, these methods ignore the benefit of multi-interest decomposition. All in all, how to model multiple interests in a multi-grained manner is the problem we want to solve.

To this end, in this paper, we proposed a novel \textbf{M}ulti-\textbf{G}rained \textbf{N}eural \textbf{M}odel~(named \baby) via a marriage between multi-interest learning and graph convolutional aggregation. Specifically, \baby is developed with two major components: \textit{user-aware graph convolution} and \textit{sequential capsule network}. We introduce a learnable process to organize a user's historical items in a user-aware manner, such that the discriminative graph structure and information propagation paths are well uncovered. We then perform graph convolution to derive the item representations iteratively, in which the complex preferences in different levels can be well captured. These multi-level item representations can better reflect the user's diverse preferences. Afterwards, a novel sequential capsule network is proposed to inject the sequential patterns into the multi-interest extraction process, leading to a more precise interest learning. The recommendation is then generated in terms of the relevance between these multiple interests of different levels and the embedding of the candidate item. To summarize, the contributions of this paper are as follows,

\begin{itemize}
    \item We propose a novel neural model by exploiting the both benefits of multi-interest learning and graph convolutional aggregation for better recommendation performance. Specifically, \baby can achieve multi-grained user preference learning by integrating multi-level preference composition and multi-interest decomposition into a unified framework.

    \item We devise a learnable graph construction mechanism to achieve discriminative structure learning over complex user behaviors. Moreover, a sequential capsule network is proposed to exploit temporal information for better multi-interest extraction.

    \item We conduct extensive experiments on three large-scale datasets collected from real-world applications. The experimental results show significant performance improvements compared with the state-of-the-art technique alternatives. Further analysis is provided to demonstrate the robustness and interpretability of \baby.
\end{itemize}
\section{Related Work}

Considering both sequential modeling and multi-interest learning in recommender systems are two major areas related to our work, we therefore briefly summarize the relevant existing methods in these two areas.

\subsection{Sequential Recommendation}
Compared with the general recommendation, the scenario of sequential recommendation is different, and its main task is simplified to predict what the user prefers for a commodity pool in the future by using considering the sequential nature of the user historical behaviors. During the early phase, traditional reasoning methods are utilized, such as Markov Chain, which assumes that the next action depends on the previous action sequence. For example, Rendle et al.~\cite{rendle2010factorization} propose to combine matrix factorization with Markov Chains~(MC) to achieve better performance in sequence recommendation. And some works assume that the next action only relies on the last behavior, using first-order Markov chain~\cite{cheng2013you}. Note that these methods are not capable to capture the long-term interests of users effectively due to the limitation of the capability to simulate the dynamic changes of user preferences over time. Then, the emergence of neural networks further enhances recommender systems' ability to extracte the preference of users, so another paradigm of sequence recommendation method based on neural networks in addition to MC-based methods has gradually become the mainstream. The most basic multi-layer perceptions~(MLPs) structure extracts the non-linear correlations from user-item interactions~\cite{he2017neural}. Then a series of models~\cite{zhang2016deep, qu2016product, shan2016deep, cheng2016wide} represented by DeepFM~\cite{guo2017deepfm} are put forward. For the DeepFM model, the FM module is used for a low-order combination of features, and the deep network module is used for the high-order combination of features. By combining the two methods in parallel, the final architecture can learn low-order and high-order combination features at the same time. 
Referring to the feature extraction mechanism in texts, audios, and pictures, CNN is used to improve the model capability in sequence recommendation. The CNN architectures are also verified to be effective in this regard to a certain extent, by mapping item sequences to embedding matrices. A representative work is Caser~\cite{tang2018personalized}, which treats the use's behavior sequence as an "image" and adopts a convolutional neural network to extract user representation. Nevertheless, this mechanism ignores the sequential relations in sequence.

Compared with approaches based on DNN and CNN, RNN is able to capture dynamic time series information~\cite{wang2020cross, zhou2018personalized}. Hidasi et al.~\cite{hidasi2015session} first introduce RNN to the sequential recommendation and achieve impressive performance gain over previous methods. Due to the appearance and excellent performance of the RNN network, more and more methods based on the RNN structure are proposed. GRU4Rec~\cite{hidasi2015session} first applies Gated Recurrent Units to model the whole session for a more accurate recommendation. To quantify the different importance of past interactions on the next prediction, attention mechanism~\cite{vaswani2017attention} is adopted. Specifically, attention mechanism makes it easy to memorize various remote dependencies or focus on important parts of the input. In addition, the attention-based methods are often more interpretable~\cite{sha2017interpretable} than traditional deep learning models. There are some other works that introduce specific neural modules for particular recommendation scenarios, which are mainly based on the combination of RNNs, CNNs, and attention structure, leading to the applications of some emerging network models coming into vogue. For example, memory networks~\cite{chen2018sequential, huang2018improving}, graph neural networks~(GNN)~\cite{wu2019session, ying2018graph} that cooperate with the attention mechanism are used to extract short-term features with more consistency or adjacency consideration. SRGNN~\cite{wu2019session} regards the session history as a directed graph. In addition to considering the relationship between an item and its adjacent previous items, it also considers the relationship with other interactive items. What's more, Fan and Liu et al.~\cite{fan2021continuous} integrate the sequence information and collaboration information, use a transformer to capture the temporal relationship in the sequence, and construct a continuous-time bipartite graph. SLi\_Rec~\cite{yu2019adaptive} utilized the fine-grained temporal characteristics of interactive data in the sequence recommendation to stress the ability to modeling sequential behaviors. The recent work represented by TGSRec~\cite{fan2021continuous} combines graph and temporal information to further greatly improve the performance of the model.

In a word, most of the existing general sequential approaches are learning to get a single representation of users from an RNN and attention-based model according to the historical behaviors. And graph models, which are capable to aggregate neighbor information, have also been proved to be very effective. Nevertheless, the user history interaction sequence contains more than one discrete interest of the user, and a single vector can not fully express the user preferences. In addition, the noise in the process of graph construction and information aggregation is also an important reason to limit the performance of graph-based sequential models.

\subsection{Multi-Interest Recommendation}
For a stronger ability to learn the complex behaviors precisely, 
recently researchers consider that representing user preferences as a single vector is insufficient, more and more sequential recommendation models based on multi-interest, therefore, appear in our field of vision. Li et al.~\cite{li2020mrif} consider that users' interests are dynamic and evolve over time. A pre-trained model based on transformer structure is designed, using the item of the next time step as the label of the interest at the current time step, and then obtains the interest of each time step. The final interest representation is generated by the attentional fusion structure. Pi et al.~\cite{pi2019practice} propose MIMN system which contains modules Neural Turing Machine~(NTM), Memory Induction Unit~(MIU), etc. In the MIU module, an additional storage unit s is also included, which contains M memory slots. It is considered that each memory slot is a user interest channel. Besides, both MIND~\cite{li2019multi} and ComiRec~\cite{cen2020controllable} devise multi-interest recommendation models on the basis of CapsNet, which uses the idea of neural routing to realize interest decomposition. Note that ComiRec introduces two multi-interest extraction mechanisms including CapsNet and self-attention. At the same time, they also have good applications in the industry. The above methods are multi-interest methods based on sequence models. With the popularity of graph neural networks, the undeniable role of neighbor information has also been proved to be effective obviously. Therefore, the approach of combining graph and multi-interest has also attracted extensive attention in recent years. For example, in SURGE~\cite{chang2021sequential}, it forms dense clusters in the interest graph to distinguish users' core interests and performs cluster-aware and query-graph graph convolutional propagation to fuse users' current core interests from behavior sequences. These mentioned approaches have also been successfully applied in many recommendation applications and are rather useful and efficient in real-world application tasks.

\section{Method}\label{sec:algo}
\begin{figure*}[htbp]
\centering
\includegraphics[width=0.96\textwidth, trim=50 80 0 0]{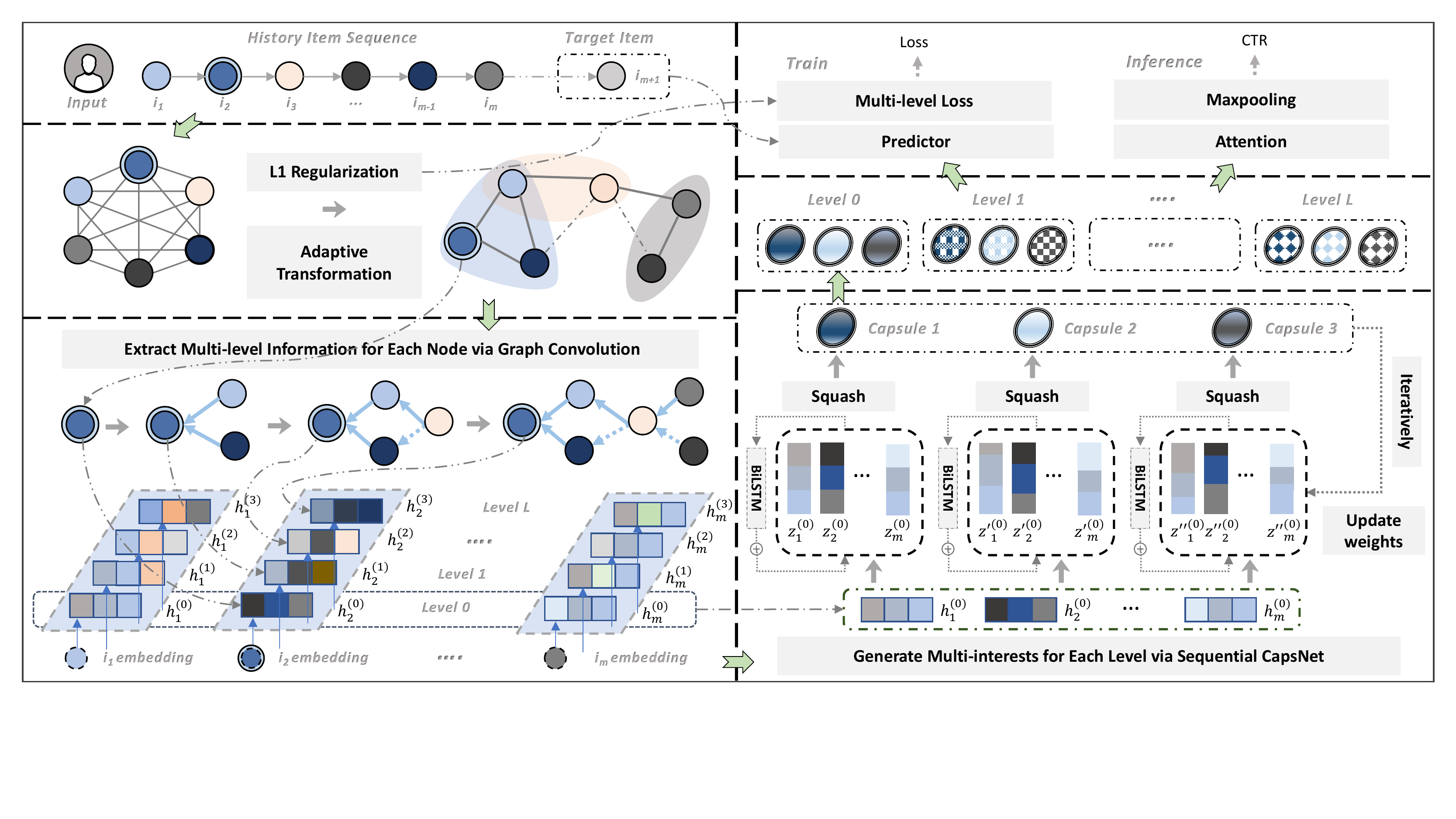}
\centering
\caption{The network architecture of our proposed \baby. The raw sequence is the historical behavior of a user. By transforming the original sequence into a user-aware adaptive graph and using the neural aggregation function of sequential CapsNet, the timing information is added to the graph in the training process. In the inference stage of the model, the max-pooling layer is used to obtain the final prediction score.}
\label{fig:model}
\end{figure*}

In this section, we present the proposed multi-grained neural model in detail. As illustrated in Figure~\ref{fig:model}, the proposed \baby consists of two main components: \textit{user-aware graph convolution} and \textit{sequential capsule network}. In the following, we firstly present the formal problem setting. Then, we describe these components, followed by the prediction and model optimization process.

\subsection{Problem Formulation}\label{ssec:formulation}

Let $\mathcal{V}=\{x_1, x_2, ..., x_M\}$ denotes the set of all items, $\mathcal{U}=\{u_1, u_2, ..., u_N\}$ be the set of all users, and $\{b_u\}_{M}^N$ be the behavior sequence set between these $N$ users and $M$ items. Here, for each user $u$, $b_u=[x_1, x_2, ..., x_m]$ is the sequence of her clicked items following the chronological order, and $m$ is the predefined maximum capacity. The sequential recommendation is to precisely identify the next item $x_{m+1}$ that user $u$ will click in terms of $\{b_u\}_{M}^N$.

\subsection{User-Aware Graph Convolution}\label{ssec:warmup}

In order to extract complex and high-order interests from user click sequences, we consider the graph structure and the aggregation of neighbor information of the target node at different distances in the graph. So at the first step, we convert discrete history behavior into a fully connected item-item graph. Compared with existing methods, we do not artificial use co-occurrence, click of the same user, and other relationships to enhance the graph, because this approach often introduces noise, which affects the performance of information aggregation in the later convolution process to some extent. In the \baby, the nodes and users embedding would be updated by using the neural aggregation of CapsNet through gradient feedback and then generate an adaptive graph structure.

\subsubsection{Embedding Layer}\label{ssec:embeddinglayer}
In the embedding layer, we firstly form a user embedding table $U\in R^{N\times d}$ and an item embedding table $V\in R^{M\times d}$, where $d$ denotes the dimension of the embedding vector. For the given user $u$ and the associated behavior sequence $b_u$, we can perform the table lookup from $U$ and $V$ to obtain the corresponding user and item embedding representations $\mathbf{x}_u$ and $[\mathbf{x}_1,\mathbf{x}_2,\cdots,\mathbf{x}_m]$ respectively. Hence, the user embeddings $U$ are expected to encode the users' overall preference, while the item embeddings $V$ reflect items' characteristics in this space instead.

\subsubsection{Graph Construction}\label{ssec:seq_trans}
Given the historical behavior sequence $b_u=[x_1, x_2, ..., x_m]$ of user $u$, we first transform the constituent items into a fully connected undirected graph $\mathcal{G}_u$ by taking each item $x_i$ as a node. It is worth mentioning that we do not condense repeated items in the sequence (\ie representing multiple clicks of the user), because the multiple clicks of the same item could convey more user preferences. We then introduce $\mathbf{A} \in R^{m\times m}$ to denote the corresponding adjacency matrix, where each entry $\mathbf{A}_{i,j}$ indicates the relatedness between item $x_i$ and item $x_j$ in the perspective of user $u$. Instead of utilizing behavior patterns to derive matrix $\mathbf{A}$, we choose to learn this relatedness based on their hidden features as follows:
\begin{align}
    \mathbf{A}_{i,j}=sigmoid((\mathbf{x}_i\odot \mathbf{x}_j) \cdot \mathbf{x}_u),
\end{align}
where $\odot$ and $\cdot$ denote the Hadamard product and inner product respectively, and $sigmoid$ denotes the activation function. We can see that the user embedding $\mathbf{x}_u$ is exploited to achieve user-aware graph construction. That is, the same item pair could have different relatedness values for different users. Also, the use of Hadamard product ensures the symmetry of the adjacency matrix.

Note that graph $\mathcal{G}_u$ is a fully connected. Hence, we need $\mathbf{A}$ to be adequately discriminative to facilitate precise multi-level preference learning. To achieve this purpose, we add $L1$ regularization on the adjacency matrix $\mathbf{A}$ to approximate a certain sparsity.

\subsubsection{Graph Convolution}
Following the common practice, we perform graph convolution operation over $\mathcal{G}_u$ as follows:
\begin{align}
    \mathbf{H}^{(l+1)} &= \delta(\Tilde{\mathbf{D}}^{-\frac{1}{2}}\Tilde{\mathbf{A}}\Tilde{\mathbf{D}}^{-\frac{1}{2}}\mathbf{H}^{(l)}\mathbf{W}),\\
    \Tilde{\mathbf{D}}^{-\frac{1}{2}} &= \mathbf{I} + \mathbf{D}^{-\frac{1}{2}}\mathbf{A}\mathbf{D}^{-\frac{1}{2}},\\
    \mathbf{H}^{(0)} &= [\mathbf{x}_1,\mathbf{x}_2,\cdots,\mathbf{x}_m]\label{eq:h0},
\end{align}
where $\mathbf{H}^{(l)}$ denotes the item representations aggregated by the $l$-th layer ($l\in \{1,\cdots,L\}$), $\delta(\cdot)$ denotes the LeakyReLU nonlinearity, $I$ denotes the identity matrix aiming to add self-loop propagation, $\mathbf{W}$ denotes the trainable parameter and $D$ denotes degree matrix of $\mathbf{A}$. The parameter matrix $\mathbf{W}$ is shared for all $L$ layers. This modification is to facilitate the feature aggregation from the high-order neighbors, which also reduces the model complexity. The item representations composited in each layer can reflect the user's diverse preferences more precisely.

\subsection{Sequential Capsule Network}\label{ssec:tsecl}
After extracting multi-level item representations $\{\mathbf{H}^{(0)},\cdots,\mathbf{H}^{(L)}\}$, where $\mathbf{H}^{l} = [\mathbf{h}_1^{(l)},\cdots,\mathbf{h}_m^{(l)}]$ and $\mathbf{h}_i^{(l)} \in R^d$, we choose to utilize CapsNet to generate the user's multiple interests for each level. Actually, the existing works for multi-interest-based recommendation utilize CapsNet to composite each interest representation through the built-in dynamic routing mechanism. The output of each capsule is equivalent to specific user interests. However, the standard dynamic routing mechanism mainly achieves the function of iterative soft-clustering. It is well validated that temporal information is critical for the sequential recommendation. This is why the application of CapsNet in fine-tuning CTR tasks is limited\cite{cen2020controllable, li2019multi}. 

Here, we repatch this defect by introducing a sequential encoding layer for CapsNet. Specifically, given the item representations at level $l$, the $i$th capsule firstly performs a linear projection over $\mathbf{H}^{(l)}$ as follows:
\begin{align}
\mathbf{Z}_i=\mathbf{H}^{(l)}\mathbf{W}_i,
\end{align}
where $\mathbf{Z}_i=[\mathbf{z}_1^{(l)},\cdots,\mathbf{z}_m^{(l)}]$ and $\mathbf{W}_i\in R^{d\times d}$ is the trainable parameter for the projection. 

We then initialize $\mathbf{g}=[g_{1},\cdots,g_m]$ by truncated normal distribution, where $g_i$ is the agreement score indicating the relevance of item $x_i$ towards the capsule. The coupling coefficient $\mathbf{c}\in R^d$ for the corresponding dynamic routing mechanism is then derived via a softmax function: 
\begin{align}
\mathbf{c}=softmax(\mathbf{g})\label{eqn:c}.
\end{align} 

Then, the capsule derive its output $\mathbf{o}_i$ via a nonlinear squashing function as follows:
\begin{align}
\mathbf{o}_i &= \frac{\Vert\mathbf{v}_i\Vert^2}{\Vert\mathbf{v}_i\Vert^2+1}\frac{\mathbf{v}_i}{\Vert\mathbf{v}_i\Vert},\\
\mathbf{v}_i &= \sum_{j=1}^m c_j\mathbf{z}_j^{(l)},
\end{align}
where $c_j$ is the $j$th element of $\mathbf{c}$. We then update the agreement score $g_i$ as follows:
\begin{align}
g_i=g_i+\mathbf{o}_i^\top\mathbf{z}_i\label{eqn:g_i}.
\end{align} 

After this first iteration, we utilize a BiLSTM\footnote{Any other sequential modeling techniques like GRU and Transformer can be straightforwardly applied here.} module to encode sequential patterns and update $\mathbf{Z}_i$ via a residual structure:
\begin{align}
\mathbf{Z}_i=\mathbf{Z}_i+BiLSTM(\mathbf{Z}_i)\label{eqn:bilstm}.
\end{align}
We then repeat the above routing process following Equation~\ref{eqn:c}-\ref{eqn:bilstm} for $\tau-1$ without further applying BiLSTM encoding over $\mathbf{Z}_i$. That is, total $\tau$ iterations are performed for each capsule. The output $\mathbf{o}_i$ in the last iteration is fed into a full-connected layer to derive the $i$th interest representation $\mathbf{q}_i^{(l)}$ in the $l$ level as follows:
\begin{align}
\mathbf{q}_i^{(l)}=ReLU(\mathbf{o}_i\mathbf{W}_i'),
\end{align} 
where $\mathbf{W}_i'\in R^{d\times d}$ is the trainable parameter. 
Assuming the number of interests is $K$, we obtain $K$ interest representations $[\mathbf{q}_1^{(l)},\cdots,\mathbf{q}_K^{(l)}]$ for the $l$th level. That is, we extract in total $(L+1)\cdot K$ interest representations.

\subsection{Prediction and Model Optimization}\label{ssec:predictor}
\subsubsection{Prediction}
Given a candidate item $x_t$, we firstly utilize an attention mechanism to derive the user preference vector $\mathbf{p}_u^{(l)}$ for $l$th level as follows:
\begin{align}
\mathbf{p}_u^{(l)} &= \sum_{j=1}^K a_j\mathbf{q}_j^{(l)},\\
a_j &= \frac{\exp(\mathbf{q}_j^{(l)\top}\mathbf{x}_t)}{\sum_{k=1}^K\exp(\mathbf{q}_k^{(l)\top}\mathbf{x}_t)},
\end{align}
where $a_j$ is the attention weight for $j$th interest. Then, we choose inner product to calculate the recommendation score as follows:
\begin{align}
\hat{y}_{u,i}^{(l)} = \mathbf{p}_u^{(l)\top}\mathbf{x}_t,
\label{eq:score}
\end{align}
where $\hat{y}_{u,i}^{(l)}$ denotes the recommendation score for the $l$th level. Note that different users could have different interest granularity. In other words, some users' interests are very complex and dynamic, the high-level user preference is more accurate. On the other hand, some users' interests are simple and straightforward, it is more appropriate to utilize the low-level user preference or even original item representations. Hence, we derive the final recommendation score by using the max-pooling: 
\begin{equation}
    \hat{y}_{u,i}=\max(\hat{y}_{u,i}^{(0)},\cdots,\hat{y}_{u,i}^{(L)}).
    \label{eq:maxpooling}
\end{equation}

\subsubsection{Model Optimization}
For the sake of enabling the model to capture user interests of different granularity from low-level to high-level, we choose to define a cross-entropy loss for each level. Thus, the final loss is formulated as follows:

\begin{align}
    \label{eq:finalloss}
    \mathcal{L}_{all} &= \sum_{l=0}^{L}\mathcal{L}_l + \theta_1\mathcal{L}_1 + \theta_2\mathcal{L}_2, \\
    \mathcal{L}_l &= -\sum_{u,i}[y_{u,i}ln(\hat{y}_{u,i}^{(l)}) + (1-y_{u,i})ln(1-\hat{u}_{u,i}^{(l)})],
\end{align}
where $y_{u,i}$ denotes the ground truth for user $u$ and item $x_i$, $\mathcal{L}_1$ denotes the $L_1$ norm of the matrix $\mathbf{A}$, $\mathcal{L}_2$ denotes the $L_2$ norm of all model parameters,  $\theta_1$ and $\theta_2$ denote the hyperparameters.

\section{Experiments}
In this section, we conduct extensive experiments on three real-world datasets from different domains for performance evaluation. We then analyze the contributions of several components and different settings for \baby\footnote{The code implementation is available at~\url{https://github.com/WHUIR/MGNM}}. Finally, a thorough analysis of ablation experiments and a framework optimizer exploration are presented.

\subsection{Experimental Settings}
\paratitle{Datasets.} The first dataset~(namely \textit{Micro-video}) is collected from a leading large-scale  Micro-video sharing platform. This dataset contains $60,813$ users and their interaction records over seven days~(\ie October 22 to October 28, 2020). We take the interactions made in the first six days as the training set. The interactions were made before $12$PM on the last day as the validation set, and the rest as the test set.

The other two datasets are from Amazon product datasets\footnote{\url{http://snap.stanford.edu/data/amazon/}}: \textit{Musical Instruments} and \textit{Toys and Games}. Here, each user interaction in the Amazon dataset is associated with a user rating score. Following the previous works~\cite{rendle2010factorizing, he2016fusing, he2017translation}, we take each user interaction with a rating score larger than $2$ as the positive. We then organize these interactions and split the interaction sequence with the ratio of $7$:$1$:$2$ to form the training, validation, and test set respectively following the chronological order. We further remove users whose length of history sequence is $1$.

Table~\ref{tab:datastats} summarizes detailed statistics of the three datasets after preprocessing. The Micro-video dataset includes a large number of items, while Toys and Games is much smaller. Also, the Musical Instruments is the smallest according to the interaction number. We can see that these three real-world datasets hold different characteristics, covering a broad range of real-world scenarios.

\begin{table}
\centering
    \caption{Statistics of the three datasets.}
    \label{tab:datastats}
    \begin{tabular}{cccc}
    \hline
      \textbf{Datasets}&\textbf{\#Users}&\textbf{\#Items}&\textbf{\#Interactions}\\
      \midrule
      Micro-video &60,813 &292,286 &14,952,659 \\
      Musical Instruments  &60,739 &56,301 &946,627 \\
      Toys and Games &313,557 &241,657 &6,212,901 \\
    \bottomrule
  \end{tabular}
\end{table}

\paratitle{Baselines.}
We compare the proposed \baby against the following state-of-the-art sequential recommendation methods:
\begin{itemize}
    \item \textbf{Caser}~\cite{tang2018personalized} is A CNN-based model which applies horizontal and vertical convolutional filters to capture the point-level, union-level, and skipping patterns for sequential recommendation.
    \item \textbf{A2SVD}~\cite{yu2019adaptive} is short for the asynchronous SVD method, which modifies the prediction model to express the user as the superposition of items. Combined with implicit feedback data, the parameters of the model are reduced, and the interpretability of the original SVD model is enhanced.
    \item \textbf{GRU2Rec}~\cite{hidasi2015session} utilizes the gated recurrent unit to model the session sequence for recommendation.
    \item \textbf{SLi\_Rec}~\cite{yu2019adaptive}  improve the traditional RNN structure by proposing a temporal-aware controller and a content-aware controller so that contextual information can guide the state transition. An attention-based framework is proposed to combine the user's long-term and short-term preferences. Hence, the representations can be generated adaptively according to the specific context.
    \item \textbf{MIMN}~\cite{pi2019practice} is a state-of-the-art multi-interest model that utilizes a multi-channel memory metwork, to capture user interests from the sequential behaviors .
    \item \textbf{MIND}~\cite{li2019multi} is a multi-interest learning model that utilizes CapsNet to capture diverse interests of a user.
    \item \textbf{ComiRec}~\cite{cen2020controllable} is a recent multi-interest model containing a multi-interest module and an aggregation module. The multi-interest module captures a variety of interests from the user behavior sequence, and can retrieve the candidate item set in a large-scale item pool. Then the aggregation module uses controllable factors to balance the accuracy and diversity for recommendation. Two variants of ComiRec are used for performance comparison: ComiRec-DR and ComiRec-SA, where CapsNet and self-attention are used for multi-interest extraction respectively.
    \item \textbf{SURGE}~\cite{chang2021sequential} is a uptodate graph neural model for sequential recommendation, which performs cluster-aware and query-graph propagation to fuse users' current core interests from behavior sequences. 
    \item \textbf{TGSRec}~\cite{fan2021continuous} is also a uptodate graph neural model that considers temporal dynamics inside the sequential patterns. 
\end{itemize}
All these baselines can be divided into four categories: \textbf{(1)} traditional sequential models that utilize RNN and attention mechanism~(\ie Caser, A2SVD and GRU4Rec); \textbf{(2)} temporal-aware models that exploit the timestamp information~(\ie SLi\_Rec and TGSRec); \textbf{(3)} multi-interest models that derive various user interest~(\ie MIMN, MIND, ComiRec-DR, ComiRec-SA and SURGE); \textbf{(4)} graph nerual models that exploit high-order correlations~(\ie SURGE). 

\begin{table*}[!]
\small
  \centering
  \caption{Performance comparison of different methods across the three datasets. The best and second-best results are highlighted in boldface and underlined respectively. $*$ indicates that the performance difference against the best result is statistically significant at $0.05$ level. Note that TGSRec took too long to train hence has no results on the large Micro-video dataset. See context for details.}
  \label{tab:amazon}
    \begin{tabular}{ccccccccccccc}
    \toprule
    \multirow{2}{*}{Method}&
    \multicolumn{4}{c}{Micro-video}&\multicolumn{4}{c}{Toys and Games}&\multicolumn{4}{c}{Music Instruments}\cr
    \cmidrule(lr){2-5} \cmidrule(lr){6-9} \cmidrule(lr){10-13}
    &GAUC&NDCG@5&HIT@5&MRR@5&GAUC&NDCG@5&HIT@5&MRR@5&GAUC&NDCG@5&HIT@5&MRR@5 \cr
    \midrule
    Caser & 0.6917* & 0.0964* & 0.1417* & 0.0815*  & 0.6234* & 0.0679* & 0.1012*  & 0.0569*& 0.6763* & 0.0955* & 0.1178* & 0.0883*\cr
    A2svd & 0.6808* & 0.0443* & 0.0686* & 0.0364*  & 0.6846* & 0.0507* & 0.0739* & 0.0430* & 0.6652* & 0.0956* & 0.1368* & 0.0820*   \cr
    GRU4Rec & 0.6944* & 0.0702* & 0.1050* & 0.0589*  &0.6624* & 0.0840* & 0.1278* & 0.0697* & 0.6498* & 0.0619* & 0.1049* & 0.0478*  \cr
    \midrule
    SLi\_rec   &  0.6903* & 0.0948* & 0.1390* & 0.0802*  & 0.7847* & 0.0932* & 0.1327* & 0.0803* &  0.6912* & \textbf{0.1078} & 0.1507* & 0.0937*      \cr
    TGSRec   & -- &-- & -- & --     & \underline{0.7915}* & \underline{0.1410}* & \underline{0.2027}* & \underline{0.1164}* & \textbf{0.7759} & 0.0946* & \underline{0.1653} & 0.0729*     \cr
    \midrule
    MIMN   & 0.7387* & \underline{0.1151}* & 0.1683* & \underline{0.0977}*  & 0.7224* & 0.1158* & 0.1676* & 0.0988* & 0.6787* & 0.0955* & 0.1509* & 0.0750*      \cr
    MIND   & 0.6778* & 0.08582* & 0.1367* & 0.0700*  & 0.6611* & 0.1015* & 0.1510* & 0.0824*& 0.6588* & 0.1040* & 0.1422* & 0.0898*      \cr   
    ComiRec-DR   & 0.7028* & 0.0863* & 0.1307* & 0.0718*  & 0.6681* & 0.1131* & 0.1597* & 0.0978*& 0.6647* & 0.1091* & 0.1541* & \underline{0.0943}*      \cr
    ComiRec-SA   & 0.6249* & 0.0354* & 0.0577* & 0.0281*  & 0.6486* & 0.0665* & 0.0977* & 0.0563* & 0.6559* & 0.0820* & 0.1204* & 0.0694*      \cr
    SURGE   & \underline{0.8116}* & 0.1091* & \underline{0.1728}* & 0.0883*   & 0.7863* & 0.0930* & 0.1353* & 0.0791* & 0.6902* & 0.1056* & 0.1494* & 0.0913*      \cr
    \midrule
    \baby &\textbf{0.8325}&\textbf{0.1463}&\textbf{0.2163}&\textbf{0.1232} & \textbf{0.8078} & \textbf{0.1611} & \textbf{0.2231} & \textbf{0.1408} & \underline{0.7480}* & \underline{0.1057} & \textbf{0.1658} & \textbf{0.1021} \cr
    \bottomrule
    \end{tabular}
\end{table*}

\paratitle{Hyperparameter Settings.}
For a fair comparison, all methods are implemented in Tensorflow and learnt with Adam optimizer. The learning rate, mini-batch size are set to $1e-3$ and $256$. The number of negative samples is $5$ in the training stage for all three datasets. We tuned the parameters of all methods over the validation and set the embedding size as $16$ and $40$ for Amazon datasets and Micro-video datasets respectively. Specifically, as to \baby, we found our model performs relatively stable when $K=4$, $L=3$, and $\theta_1=1e-6, \theta_2=1e-5$.

\paratitle{Evaluation Metric.}
Following the same setting in~\cite{fan2021continuous}, we sample $1,000$ negative items for each testing instance. Four common metrics: hit rate~(HR), mean reciprocal rank~(MRR), and normalized discounted cumulative gain~(NDCG) and Group AUC~(GAUC), are used for performance evaluation. For method, we repeat the experiment $5$ times and report the average results. The statistical significance test is conducted by the student's $t$-$test$.

\subsection{Performance Evaluation}
The overall performance of all methods is reported in Table~\ref{tab:amazon}. Here, we make the following observations.

As for traditional sequential models that utilize RNN and attention mechanism, they are difficult to achieve better performance. Compared with temporal-aware models and multi-interest models, they are not suitable for complex and various user interest modeling. The temporal-aware models perform very well in Amazon datasets. Specifically, on the Music Instruments dataset, TGSRec and Sli\_Rec achieve the best performance in terms of GAUC and NDCG@$5$ respectively. They also achieve strong performance in terms of the other six metrics for both Toys and Games and Music Instruments. It is worthwhile to mention that TGSRec needs to build a global graph and takes the interactions at different time points as edges. This design choice requires much more computation cost for graph retrieval and convolution. Note that the graph constructed on the Micro-video dataset contains more than $200$ million edges. We utilize the implementation released by the original authors for evaluation. The time of an epoch training exceeds $1,200$ hours. Hence, we do not obtain results on the Micro-video dataset. 

Also, given the superiority of these temporal-aware models on both Amazon datasets, the multi-interest models perform better on the Micro-video dataset. This suggests that neither temporal-aware models nor multi-interest models are robust enough to achieve precise preference understanding across different scenarios. Considering the semantic space in the Micro-video recommender scenario could be much broader than commodities in E-Commerce scenarios, the interest of each user also becomes more complex. It is reasonable that the multi-interest models could achieve better recommendation performance instead.

Our proposed \baby has obvious improvement in most settings for the three datasets including Micro-video, Toys and Games, and Music Instruments. In detail, \baby performs significantly better than all the baselines on $10$ out of $12$ dataset and metric combinations.  Although our \baby achieves only comparable NDCG@$5$ performance against SLi\_Rec and performs a bit worse to TGSRec in terms of GAUC both on the Music Instruments dataset, we need to emphasize that both SLi\_Rec and TGSRec exploit additional timestamp features to seize more discriminative capacity. In other words, our model \baby lacks one-dimensional timestamp characteristics than these two models~(i.e. SLi\_Rec and TGSRec). Note that modeling multi-grained multi-interest in \baby and exploiting timestamp information are not mutually excluded. As shown in Equation~\ref{eqn:bilstm}, it is straightforward to include fine-grained timestamp information in the sequential capsule network component\footnote{We leave it as a part of our future work.}. Moreover, we can see that our proposed \baby performs increasingly better on larger datasets. The relative performance gain by \baby against the best baseline is in the range of $1.63\%-2.44\%$ and $2.09\%-4.35\%$ on Toys and Games and Micro-video datasets respectively. This further confirms that our \baby is effective to capture multi-grained user interests for large-scale real-world scenarios that are rich in semantics.


\subsection{Model Analysis}
Here, we investigate the impact of each design choice and important parameter settings to the performance of \baby.

\eat{
\begin{table}[htbp]
  \centering
  \caption{The Impact of the TSECL Components on Toys and Games Dataset.}
  \label{tab:tescl}
    \begin{tabular}{ccccc}
    \toprule
    \multirow{2}{*}{Model}&
    \multicolumn{4}{c}{Toys and Games}\cr
    \cmidrule(lr){2-5}
    &GAUC&NDCG@5&HIT@5&MRR@5 \cr
    \midrule
    SCN$\to$ SelfAtt & 0.6724 & 0.0791 & 0.1148 & 0.0674  \cr
    SCN$\to$ SumPooling & 0.6651 & 0.0846 & 0.1232 & 0.0720  \cr
    SCN$\to$ BiLSTM & 0.6589 & 0.0838 & 0.1223 & 0.0712  \cr
    SCN (Transformer) & 0.6663 & 0.0923 & 0.1321 & 0.0792  \cr
    TSECL & 0.8078 & 0.1611 & 0.2231  & 0.1408  \cr
    \bottomrule
    \end{tabular}
\end{table}
}

\begin{figure*}[!] 
    \centering
    \begin{subfigure}[t]{4.3cm}
        \centering
        \includegraphics[width=4.3cm]{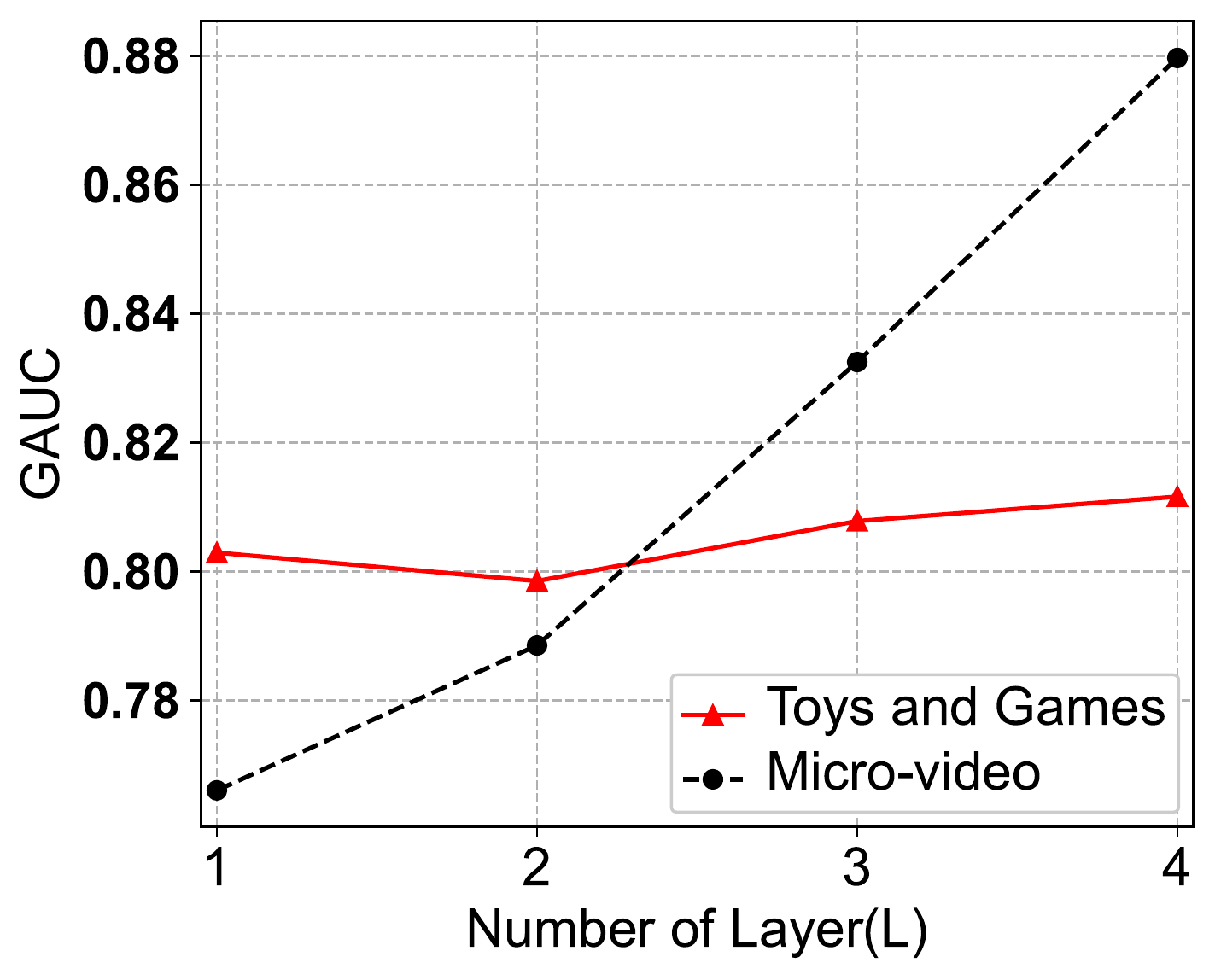}
        \caption{GAUC}      
    \end{subfigure}
    \quad
    \begin{subfigure}[t]{4.3cm}
        \centering
        \includegraphics[width=4.3cm]{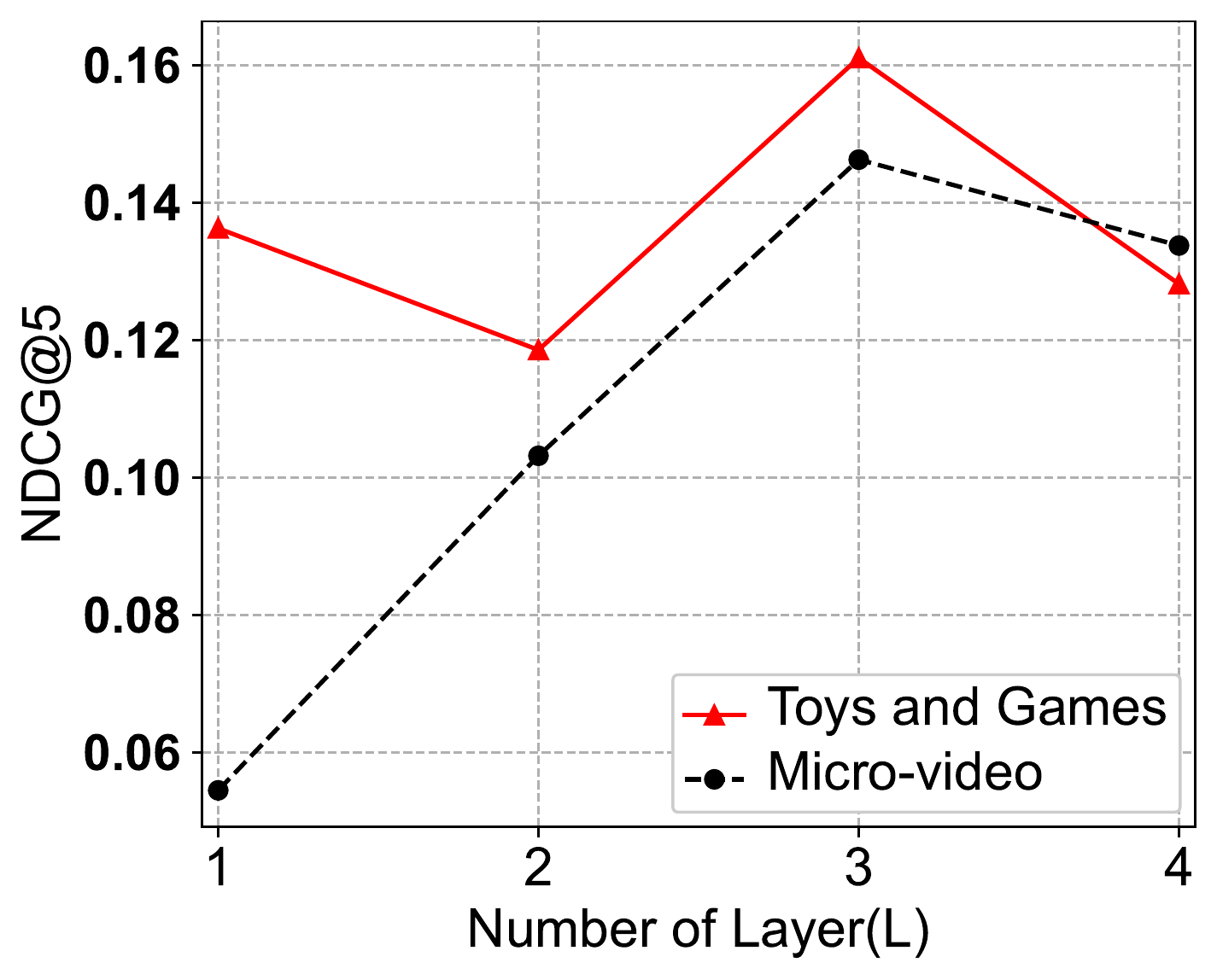}
        \caption{NDCG@5}
    \end{subfigure}
    \begin{subfigure}[t]{4.3cm}
        \centering
        \includegraphics[width=4.3cm]{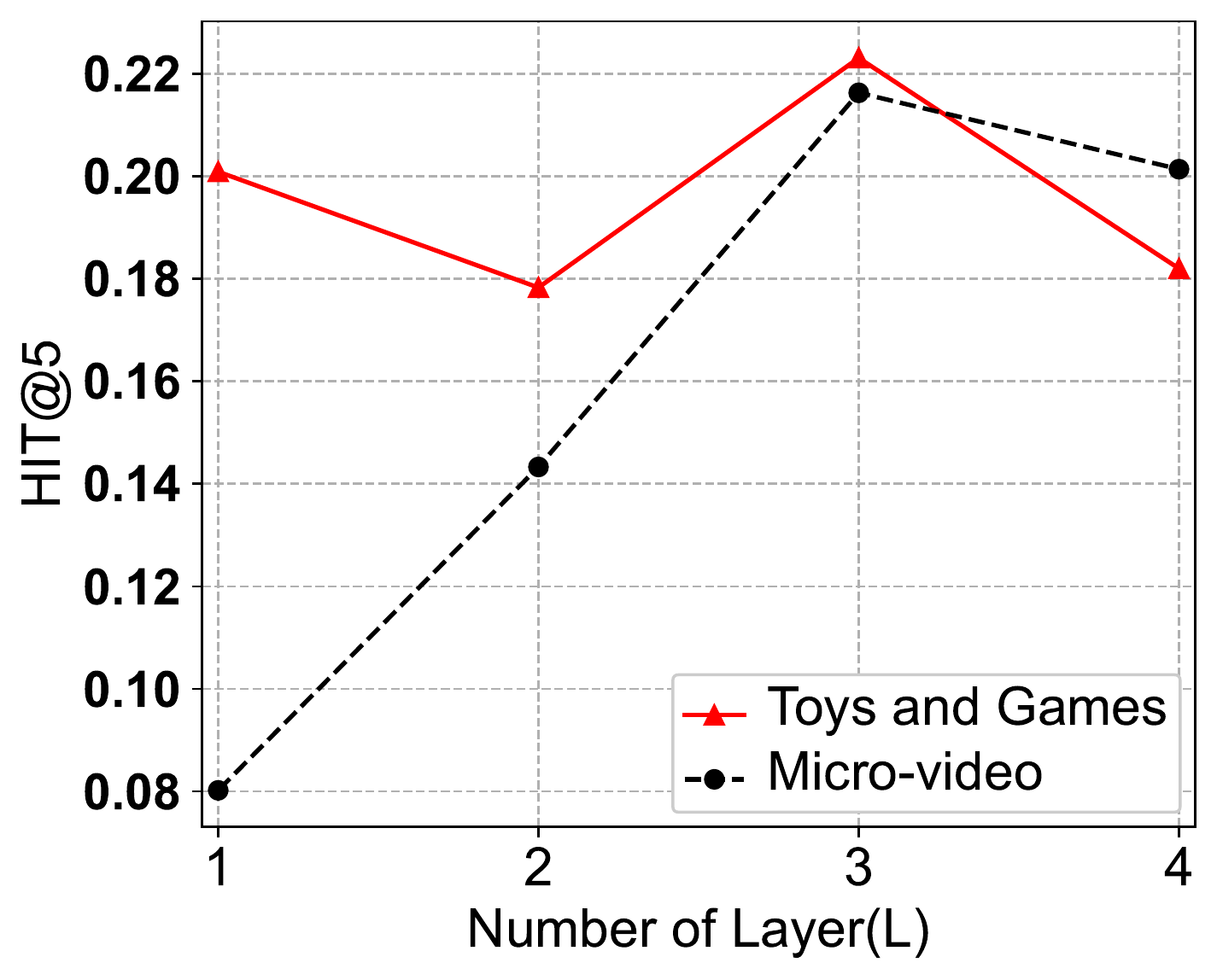}
        \caption{HIT@5}
    \end{subfigure}
    \begin{subfigure}[t]{4.3cm}
        \centering
        \includegraphics[width=4.3cm]{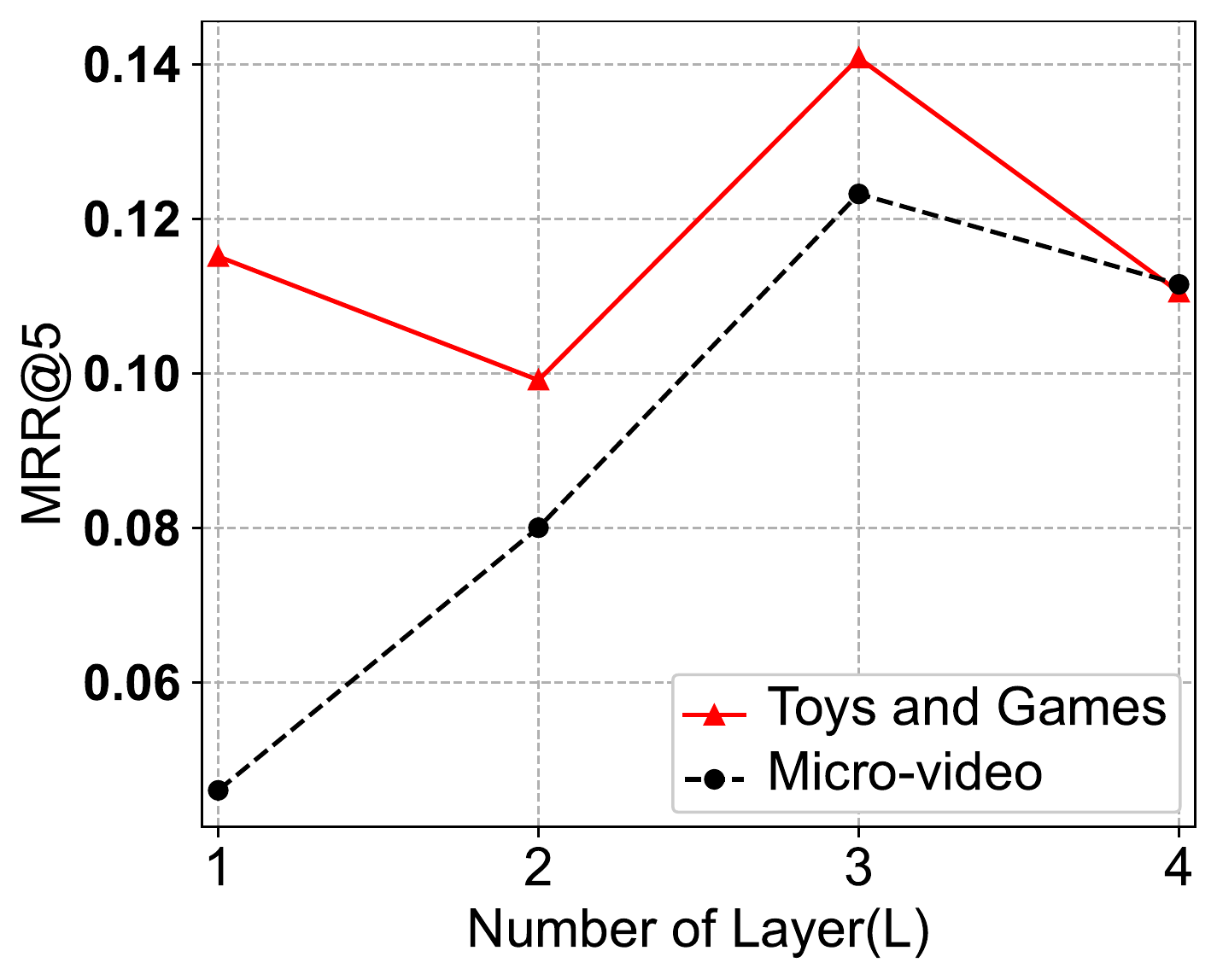}
        \caption{MRR@5}
    \end{subfigure}
    \caption{The performance of different $L$ values on Toys and Games and Micro-video Datasets.}\label{fig:layers}
\end{figure*}

\begin{figure*}[!] 
    \centering
    \begin{subfigure}[t]{4.3cm}
        \centering
        \includegraphics[width=4.3cm]{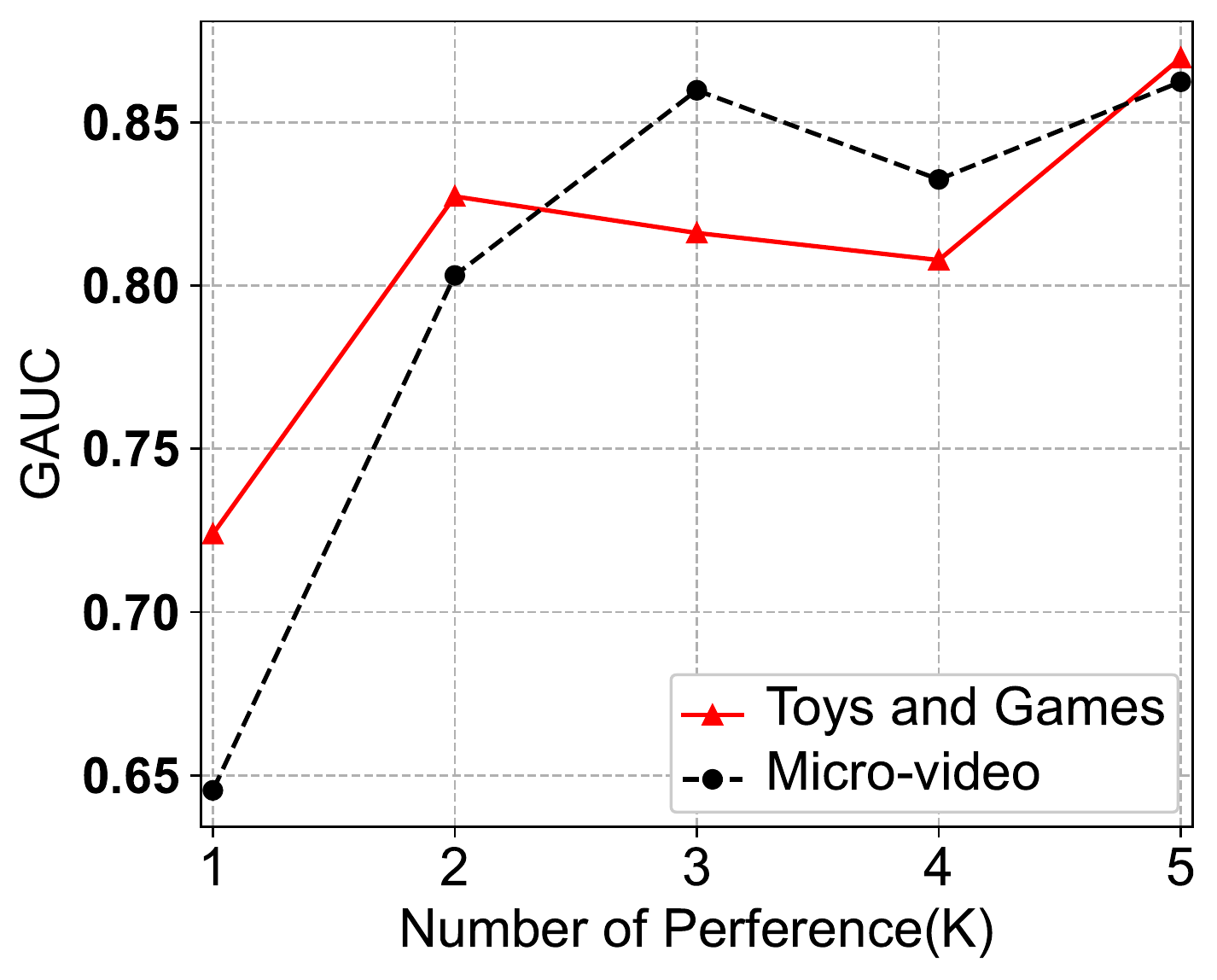}
        \caption{GAUC}      
    \end{subfigure}
    \quad
    \begin{subfigure}[t]{4.3cm}
        \centering
        \includegraphics[width=4.3cm]{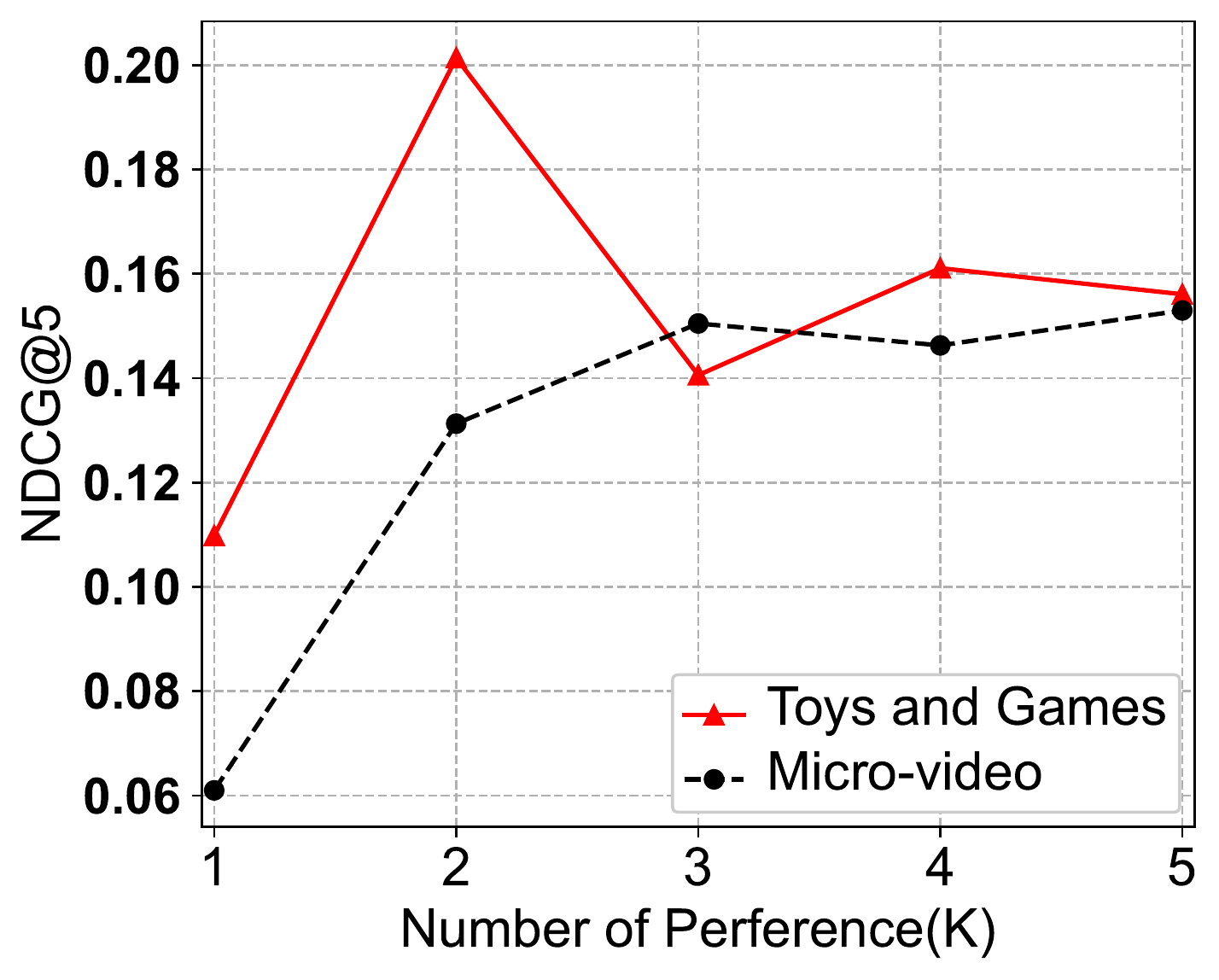}
        \caption{NDCG@5}
    \end{subfigure}
    \begin{subfigure}[t]{4.3cm}
        \centering
        \includegraphics[width=4.3cm]{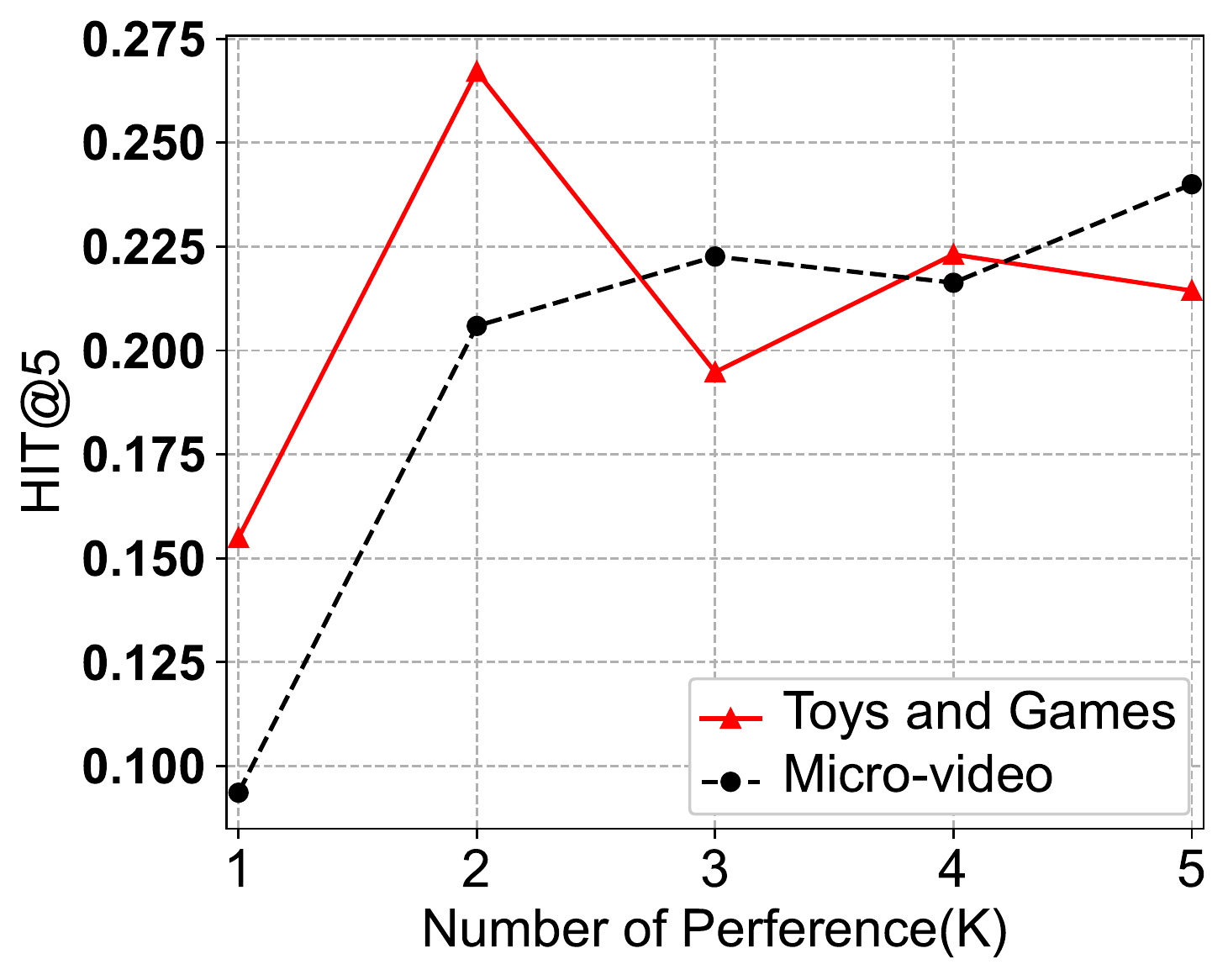}
        \caption{HIT@5}
    \end{subfigure}
    \begin{subfigure}[t]{4.3cm}
        \centering
        \includegraphics[width=4.3cm]{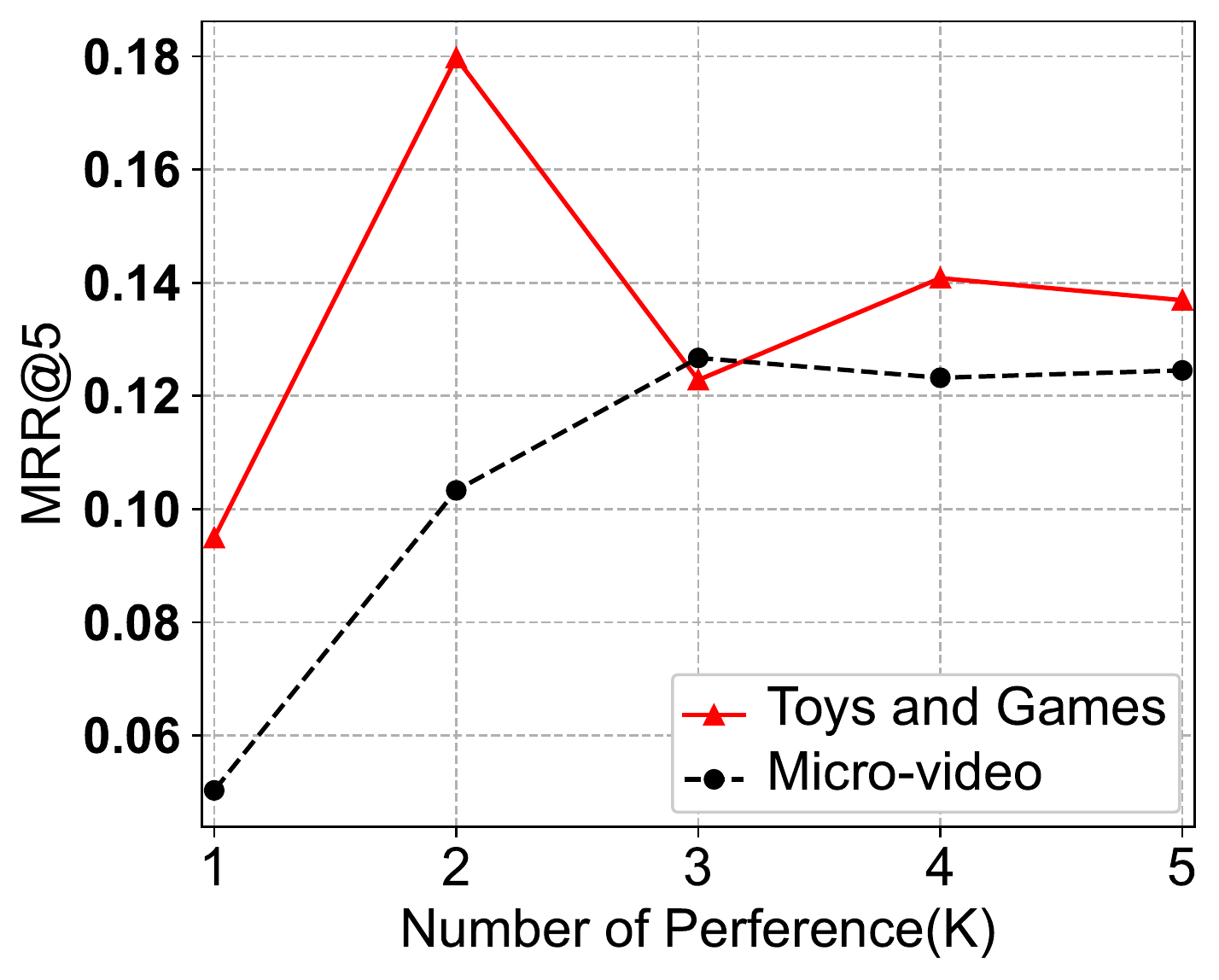}
        \caption{MRR@5}
    \end{subfigure}
    \caption{The performance of different $K$ values on Toys and Games and Micro-video Datasets.}\label{fig:preferences}
\end{figure*}

\paratitle{Ablation Study.} We conduct an ablation study for each design choice in \baby to justify their validity. Specifically, these factors include user-aware graph convolution~(UGCN), the $L1$ regularization on the adjacency matrix $A$~($L1$Norm), sequential capsule network without sequential encoding layer~(BiLSTM), and the max-pooling based prediction~(MaxPool). As to the sequential capsule network, we also examine the following variants:

\begin{itemize}
    \item SCN$\to$ BiLSTM: We replace the sequential capsule network with BiLSTM to encode the user behavior sequence in different levels. The last hidden state generated by the BiLSTM is taken the user interest in the corresponding level.
    \item SCN$\to$ SumPool: We replace the sequential capsule network with a sum pooling mechanism. Similar to SCN$\to$ BiLSTM, the resultant representation is taken the user interest for the corresponding level.
    \item SCN$\to$ SelfAtt: We replace the sequential capsule network with a self-attention mechanism. The candidate item is utilized to derive the user interest for each level by using an attention mechanism.
    \item SCN (Transformer): We replace the built-in BiLSTM in the sequential capsule network with a powerful transformer module. 
\end{itemize}

Table~\ref{tab:ablation} reports the performance of these variants and the full \baby model on Toys and Games dataset\footnote{Similar observations are also made in the other two datasets}. Here, we can make the following observations.

Firstly, The $L_1$ regularization indeed improves the discriminative capacity of user-aware graph convolution. The experimental results show that the performance degradation without it is obvious. When we remove the multi-level item representation learning supported by user-aware graph convolution~(\ie w/o UGCN), substantial performance degradation is also experienced by \baby, which illustrates the effectiveness of user-aware graph convolution and multi-level preference learning significantly. Also, we can find that \baby experiences a large performance degradation by removing the sequential encoding layer~(\ie w/o BiLSTM). This is reasonable since the sequential patterns have been well validated to be effective for the sequential recommendation. Now, we further validate that sequential patterns are also very useful for multi-interest learning. At last, the max-pooling-based prediction plays a great role in improving all four performance metrics. As we described earlier, the max-pooling mechanism is flexible in capturing the complex user preference from multi-grained interests.

\begin{table}[!]
  \centering
  \caption{The ablation study of \baby on Toys and Games Dataset. The best results are highlighted in boldface.}
  \label{tab:ablation}
    \begin{tabular}{ccccc}
    \toprule
    \multirow{2}{*}{Model}&
    \multicolumn{4}{c}{Toys and Games}\cr
    \cmidrule(lr){2-5}
    &GAUC&NDCG@5&HIT@5&MRR@5 \cr
    \midrule
    w/o UGCN & 0.7499 & 0.0929 & 0.1325 & 0.0799 \cr
    w/o $L1$Norm & 0.7757 & 0.1306 & 0.1848 & 0.1128  \cr
    w/o BiLSTM & 0.6743 & 0.1205 & 0.1689 & 0.1046  \cr
    w/o MaxPool & 0.8491 & 0.0980 & 0.1430 & 0.0832 \cr
    \midrule
    SCN$\to$ BiLSTM & 0.6589 & 0.0838 & 0.1223 & 0.0712  \cr
    SCN$\to$ SumPool & 0.6651 & 0.0846 & 0.1232 & 0.0720  \cr
    SCN$\to$ SelfAtt & 0.6724 & 0.0791 & 0.1148 & 0.0674  \cr
    SCN (Transformer) & 0.6663 & 0.0923 & 0.1321 & 0.0792  \cr
    \midrule
    \baby & \textbf{0.8078} & \textbf{0.1611} & \textbf{0.2231}  & \textbf{0.1408}  \cr
    \bottomrule
    \end{tabular}
\end{table}

Secondly, we further dive deep into the effectiveness of the sequential capsule network component. The first three variants~(\ie SCN$\to$ BiLSTM, SCN$\to$ SumPool, SCN$\to$ SelfAtt) aim to remove the multi-interest modeling by considering only the multi-level user preferences. We can see that these three variants all experience significant performance degradation across the four metrics. Recall that \baby w/o UGCN also produces a substantial performance degradation above. These two observations suggest that both user-aware graph convolution and sequential capsule network works as a whole and either of them complements the other, leading to better user preference understanding. At last, we find that encoding sequential patterns with a heavy module like transformer achieves better performance than SCN$\to$ BiLSTM, SCN$\to$ SumPool, and SCN$\to$ SelfAtt in terms of NDCG@$5$, HIT@$5$ and MRR@$5$. This also validates the benefit of modeling sequential patterns for multi-interest learning. However, the huge number of parameters involved in a transformer module could complicate the model optimization process. Note that we also derive multi-level item representations by the user-aware graph convolution, a lightweight sequential model like BiLSTM is sufficient for the next step. At the same time, to the best of our knowledge, there are no previous methods to integrate sequential modeling with CapsNet.

\paratitle{Impact of $L$ Value.}
Recall that we stack $L$ layers of graph convolution in \baby to reflect the user's diverse preferences in multi-grained manner. A larger $L$ value can recruit more high-order neighbors to derive the user's preference more and more distant neighbor information is aggregated. However, some noisy information would also be included to deliver adverse impact. In Figure~\ref{fig:layers}, we plot the performance patterns of varying $L$ values for both Toys and Games and Micro-video datasets. It is reasonable to see that all NDCG@$5$, HIT$5$ and MRR@$5$ scores firstly increase when $L$ becomes large~($L\leq 3$), and then decrease when $L$ is too large~($L>3$). Also, the metric GAUC seems to be very stable for different $L$ values.

\paratitle{Impact of $K$ Value.}
The number of interests $K$ in \baby controls the diversity of user preferences. Figure~\ref{fig:preferences} plots the performance patterns of varying $K$ values for both Toys and Games and Micro-video datasets. We can observe that a single interest representation~(\ie $K=1$) achieves the worst performance across the four metrics. The optimal $K$ value is $2$ and $4$ for Toys and Games and Micro-video datasets respectively. Moreover, we can see that \baby achieves relatively more stable performance when $K$ is in the range of $[3,5]$. This is reasonable since the semantic space of the Micro-video dataset is much broader than the Toys and Games dataset.

\begin{figure}[!]
\center
\includegraphics[width=0.98\linewidth]{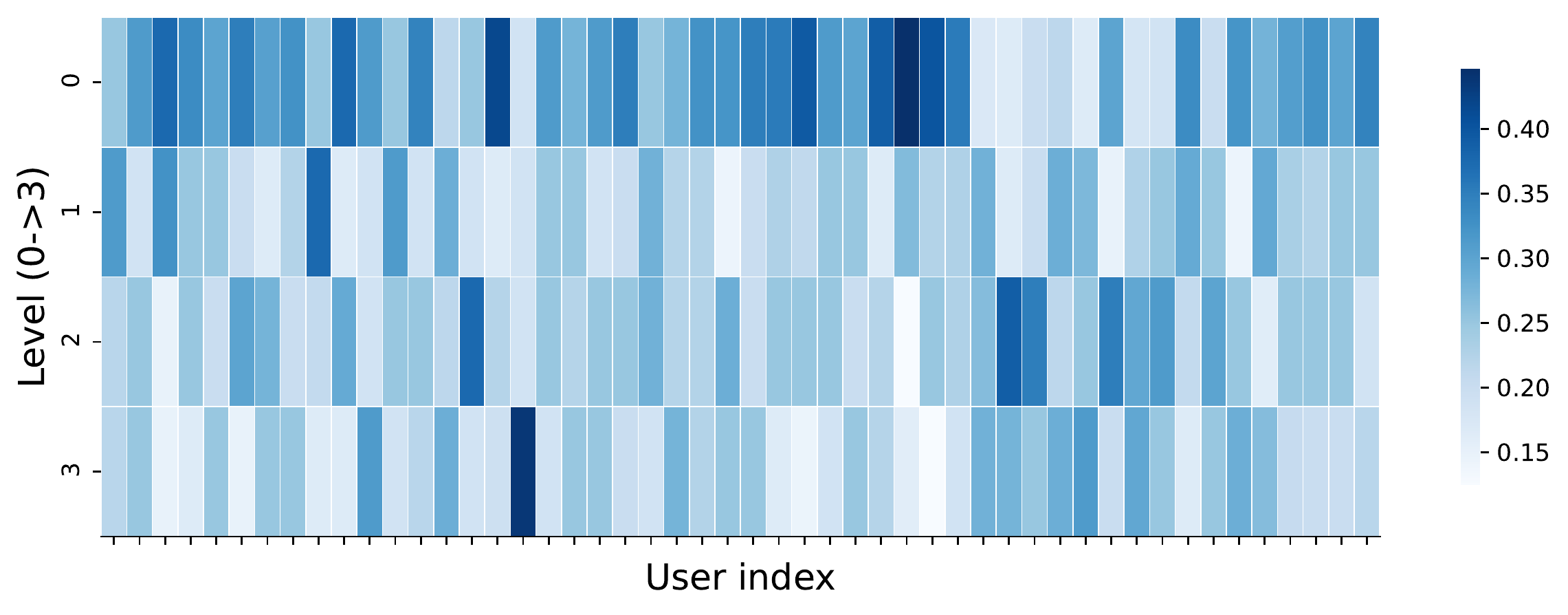}
\caption{Visualization of multi-level user interest distribution on Micro-video dataset (\textit{Best viewed in color}).}
\label{fig:vis_mv}
\end{figure}

\begin{figure}[!]
\center
\includegraphics[width=0.98\linewidth]{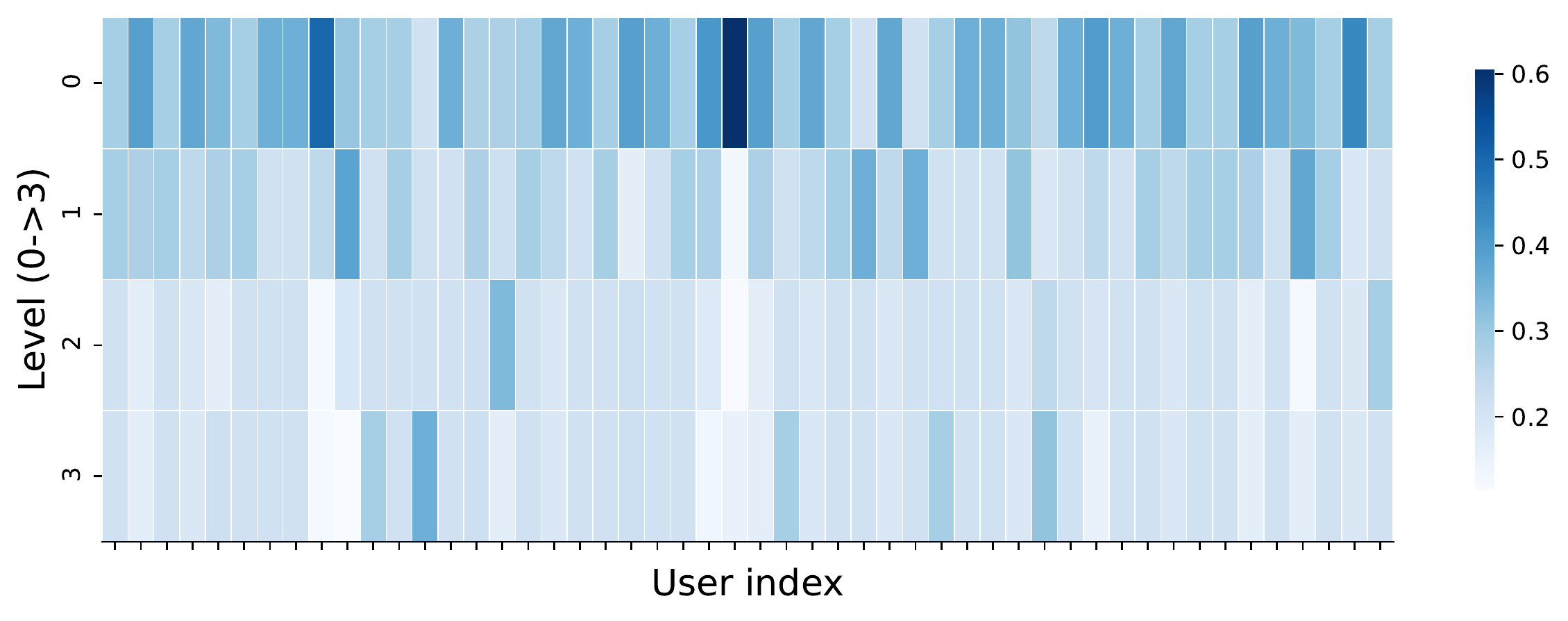}
\caption{Visualization of multi-level user interest distribution on Toys and Games dataset (\textit{Best viewed in color}).}
\label{fig:vis_tg}
\end{figure}

\begin{figure}[!]
\center
\includegraphics[width=0.98\linewidth]{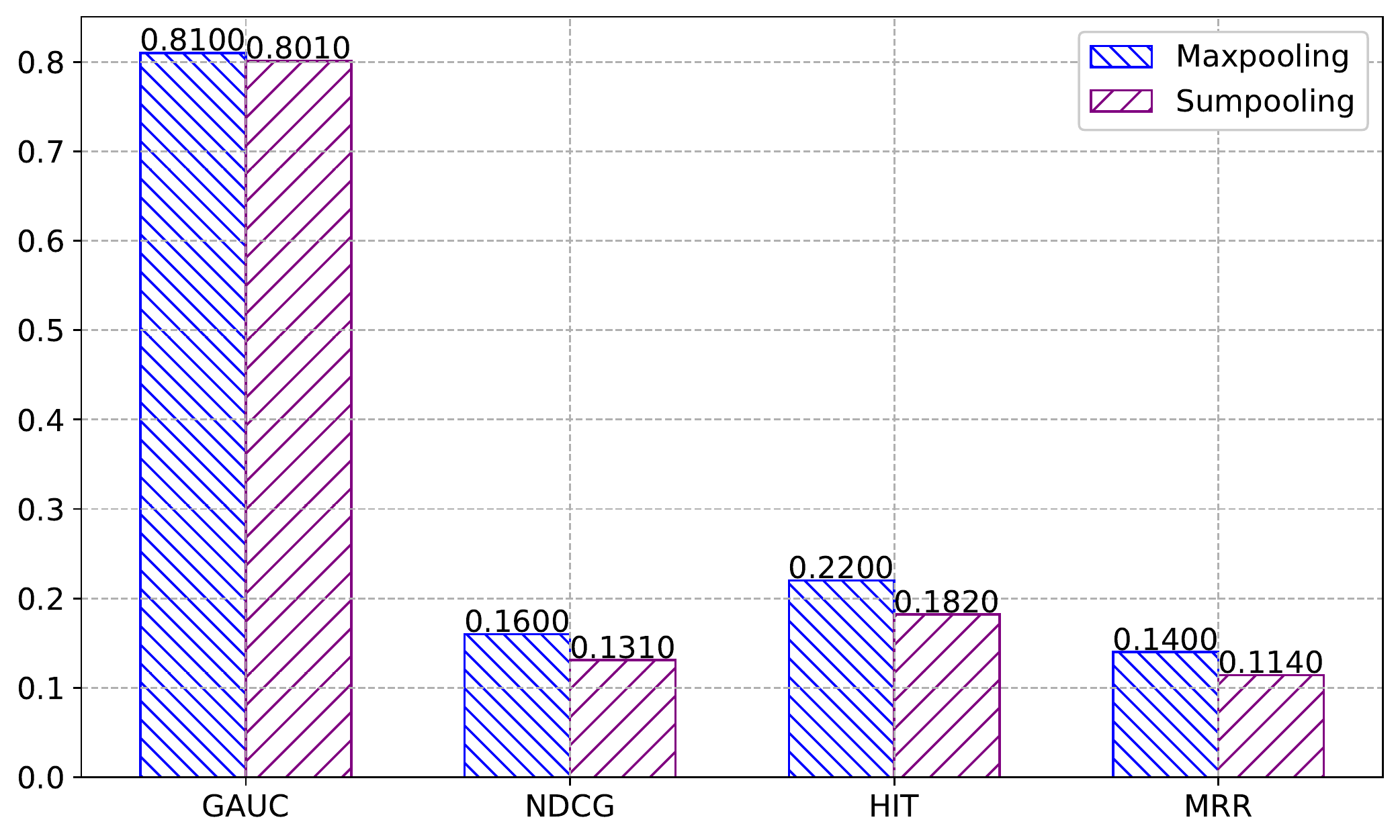}
\caption{Max-pooling vs. sum-pooling for \baby in the inference stage.}
\label{fig:max_vs_sum}
\end{figure}

\paratitle{Multi-Level User Interest Distribution.}
In Figure~\ref{fig:layers}, we examine the impact of different $L$ layers in Micro-video. Here, we further investigate whether the multi-level user preferences indeed perform different roles for different users. Specifically, we randomly sample $50$ users from the Micro-video and Toys and Games datasets, respectively. For each user, we include her positive items in the test set and thousands of random negative items, and count the activated preference level by the max-pooling based predictor~(ref. Equation~\ref{eq:maxpooling}). 
Figure~\ref{fig:vis_mv} and Figure~\ref{fig:vis_tg} plots the distribution of these activated levels for each user on Micro-video and Toys and Games datasets, respectively. We can observe that the desired preference level is quite different for different users. Also, the first two layers are adequate for most users in \baby on Toys and Games dataset. But we also need to derive high-level preferences for a few users~(\ie $L\ge 2$) in Figure~\ref{fig:vis_tg}. As for the Micro-video dataset with a larger semantic space, the role of high-level preferences becomes more significant to all users from Figure~\ref{fig:vis_mv}. On the whole, users have more high-level preferences on Micro-video. In other words, users' interests in Micro-video scenes are higher-level, more complex, and change faster, which we mentioned in figure~\ref{fig:hist} and the previous analysis. Thus, this phenomenon is in line with our expectations, which well proves the effective impact of the multi-level mechanism. Furthermore, in the inference stage, we replace max-pooling with sum-pooling to further verify the influence of max-pooling structure in Figure~\ref{fig:max_vs_sum}. 
In combination with the distribution of user interests in Figure~\ref{fig:vis_mv} and~\ref{fig:vis_tg}) and the better experimental results of max-pooling than sum-pooling in Figure~\ref{fig:max_vs_sum}, this suggests that the user-aware graph convolution does distinguish the user's interest in multi-level precisely and the multi-level mechanism does promote the performance.








\begin{table}[!]
  \centering
  \caption{Runtime comparisons for different datasets.}
  \label{tab:time}
    \begin{tabular}{cccc}
    \toprule
    Datasets&Per Iteration (s)&Iterations&Total Time (m) \cr
    \midrule
    Micro-video & 0.3825 & 15,311 & 97.60 \cr
    Toys and Games & 0.1843 & 13,202 & 40.55 \cr
    Music Instruments & 0.0598 & 2,373 & 2.37  \cr
    \bottomrule
    \end{tabular}
\end{table}

\paratitle{Time Complexity Analysis.} Table~\ref{tab:time} reports the runtime of \baby training procedure for a single user on different datasets by using a single GPU. Although the \baby adopt the graph convolution, we can see that the model training with $15$M interactinos  takes about $1.5$H for one epoch, which is computationally efficient.

\section{Conclusion }
In this paper, we proposed a novel \textbf{m}ulti-\textbf{g}rained \textbf{n}eural \textbf{m}odel~(named MGNM) with a combination of multi-level and multi-interest as a unified solution for sequential recommendation task.
A learnable process was introduced to re-construct loose item sequences into tight item-item interest graphs in a user-aware manner. We then performed graph convolution to derive the item representations iteratively, in which the complex preferences in different levels can be well captured. Afterwards, a novel sequential CapsNet was designed to inject the sequential patterns into the multi-interest extraction process, leading to a more precise interest modeling. Extensive experiments on three real-world datasets in different recommendation scenes demonstrated the effectiveness of the multi-level and multi-interest mechanisms. Further studies on the number of preferences and multi-level user interest distribution confirmed that our method was able to deliver recommendation interpretation at multi-level granularities.

\begin{acks}
    This work was supported by National Natural Science Foundation of China (No.~61872278); and Young Top-notch Talent Cultivation Program of Hubei Province. Chenliang Li is the corresponding author.
\end{acks}

\balance
\bibliographystyle{ACM-Reference-Format}
\bibliography{refer}

\end{document}